Comment: Latex file, 10 pages, no figures.
\documentstyle[12pt]{article}
\begin{document}
\baselineskip 24pt
\setlength{\textheight}{8.5in}
\setlength{\topmargin}{0in}
\begin{centering}
\LARGE {\bf A Quantum Bit Commitment Protocol Based on EPR States}
\\  \vspace{.75in}
\Large {M. Ardehali} \footnote[1] {email:ardehali@mel.cl.nec.co.jp}
\,$^,$\footnote[2]
{Atago, Tama-shi, Tokyo 206 Japan;
permanent address:
Microelectronics Research Laboratories,
NEC Corporation,
Sagamihara,
Kanagawa 229
Japan
Correspondence should only be sent to permanent address.}
\\ \vspace{.75in}
\end{centering}

\begin{abstract}
A protocol for quantum bit commitment
is proposed.
The protocol is feasible with
present technology and is secure
against cheaters with unlimited computing power as long as the
sender does not have the technology
to store an EPR particle for an arbitrarily long
period of time. The protocol is very efficient,
requiring only tens of particles.
\end{abstract}
\pagebreak

Quantum cryptography was initiated by the
pioneering work of Wiesner in late Sixties (his paper, however,
was not published until 1983 \cite{1}).
Over the past two decades, many applications of
quantum cryptography have been discovered \cite{2},\cite{3}.
One of the most
outstanding applications of quantum cryptography
is quantum bit commitment.

The interest in bit commitment is
motivated by a recent trend in cryptographic research
to reduce or more preferably to eliminate the complexity
assumptions from the protocols.
Many protocol problems that were previously
solved subject to complexity
assumptions are now being solved
without these assumptions. These
breakthroughs demonstrate the weaknesses of the unproven complexity
assumptions. One way of eliminating the complexity assumptions from
a protocol is to build the protocol using a small set of
relatively simple primitives. The security of the  protocol then
entirely depends on the security of its primitives.
One of the most fundamental
primitives is the bit commitment. The extreme generality
and usefulness of bit commitment primitive
has been demonstrated by several authors \cite{4}.

Let us briefly review the goal of the bit commitment protocol:
\\
(1) Alice has a bit $\lambda$ in mind to which she would like to
be committed toward Bob.
\\
(2) Bob should not learn any information about $\lambda$ before
Alice opens up the commitment.
\\
(3) Alice should not be able to change the value of $\lambda$ after
the commitment.

In the past, several bit commitment protocols based on complexity
assumptions have been proposed.
However, none of these protocols are safe
against cheaters with unlimited computing power.
Bit commitment protocols based on uncertainty principle
have also been proposed\cite{5} \cite{6} \cite{7}.
However, all these protocols are insecure
against EPR attack [8-11].
In this paper, we describe a new and efficient
quantum bit protocol
which requires only tens of EPR particles and
is feasible
with present technology. However, the present
protocol, similar to previous schemes,
is not secure against a cheating Alice who has the technology to
store an EPR
particle for an arbitrarily long period of time.

Before proceeding, it is useful to review some elementary features
of quantum mechanics. We consider a pair of
particles in the EPR entangled state
\begin{math}
\mid\!\Phi\rangle=\frac{\displaystyle 1}
{\displaystyle \sqrt{2}}\left(\mid\uparrow
\uparrow\rangle+\mid\downarrow\downarrow\rangle\right)
\end{math}.
After particles are spatially separated, the spin of the
first (second) particle  $m_{1}^{a}$ ($m_{2}^{b}$) is measured
along an arbitrary axis $\vec{a}$  ($\vec{b}$), with
the $z$ axis being along the direction of flight
of particles. If the spin of the first (second) particle is
up, then $m_{1}^{a}=1$ ($m_{2}^{b}=1$), and if the spin
of the first (second) particle is down, then
$m_{1}^{a}=-1$ ($m_{2}^{b}=-1$).
The expected value of the product of the spins
of the particles is
\begin{eqnarray}
E_{\Phi}(\vec{a}, \vec{b})&=&
\langle\Phi\!\mid\sigma_{1}^{a}\sigma_{2}^{b}
\mid\!\Phi\rangle\\ \nonumber &=&
\cos \theta_1\cos \theta_2+ sin \theta_1\sin\theta_2
\cos\left(\phi_1+\phi_2\right),
\end{eqnarray}
where $\theta_1(\theta_2)$ is the polar angle
between $\vec{a}\left(\vec{b}
\right)$ and the $z$ axis,
and $\phi_1(\phi_2)$ is the azimuthal angle
between $\vec{a}\left(\vec{b}
\right)$ and the $x$ axis.
Similarly for a pair of particles in the EPR entangled state
\begin{math}
\mid\!\Phi' \rangle=\frac{\displaystyle 1}
{\displaystyle \sqrt{2}}\left(\mid\uparrow
\uparrow\rangle-\mid\downarrow\downarrow\rangle\right),
\end{math}
\begin{eqnarray}
E_{\Phi'}(\vec{a}, \vec{b}) &=&
\langle\Phi'\!\mid\sigma_{1}^{a}\sigma_{2}^{b}
\mid\!\Phi' \rangle \\ \nonumber
&=&\cos\theta_1\cos\theta_2 -  sin\theta_1\sin\theta_2
\cos\left(\phi_1+\phi_2\right).
\end{eqnarray}
For a pair of particles in the entangled state
\begin{math}
\mid\!\Psi \rangle=\frac{\displaystyle 1}
{\displaystyle \sqrt{2}}\left(\mid\uparrow
\uparrow\rangle+ i \mid\downarrow\downarrow\rangle\right),
\end{math}
\begin{eqnarray}
E_{\Psi}(\vec{a}, \vec{b}) &=&
\langle\Psi \!\mid\sigma_{1}^{a}\sigma_{2}^{b}
\mid\!\Psi \rangle \\ \nonumber
&=&\cos\theta_1\cos\theta_2 +  sin\theta_1\sin\theta_2
\sin\left(\phi_1+\phi_2\right).
\end{eqnarray}
\\
Finally for a pair of particles in the entangled state
\begin{math}
\mid\!\Psi' \rangle=\frac{\displaystyle 1}
{\displaystyle \sqrt{2}}\left(\mid\uparrow
\uparrow\rangle- i \mid\downarrow\downarrow\rangle\right),
\end{math}
\begin{eqnarray}
E_{\Psi'}(\vec{a}, \vec{b}) &=&
\langle\Psi'\!\mid\sigma_{1}^{a}\sigma_{2}^{b}
\mid\!\Psi'\rangle \\ \nonumber
&=&\cos\theta_1\cos\theta_2 -  sin\theta_1\sin\theta_2
\sin\left(\phi_1+\phi_2\right).
\end{eqnarray}
For states $\mid\Phi \rangle$, $\mid\Phi'\rangle$,
$\mid\Psi\rangle$, and $\mid\Psi'\rangle$, the
probability that the product of the spins of the two particles
$m_{1}^{a}m_{2}^{b}$ is $+1$ or $-1$ is
\begin{eqnarray}
p(m_{1}^{a}m_{1}^{b}=1)&=& \frac{1 +
E(\vec{a}, \vec{b})}{2},
\\ \nonumber
p(m_{1}^{a}m_{1}^{b}=-1)&=& \frac{1 -
E(\vec{a}, \vec{b})}{2}.
\end{eqnarray}

With the above in mind, we now proceed to describe a
quantum bit commitment protocol based on EPR states.
Alice and Bob initiate
the following steps:
\\
(1) Alice and Bob agree on a security parameter $n$. They also
agree that
if Alice wants to be committed to bit
$\lambda = 1$, then
she prepares a sequence of $n$ states, randomly chosen from
$\mid\!\!\Phi\rangle$ or
$\mid\!\!\Phi'\rangle$,
and if she wants to be committed to bit
$\lambda = 0$, then 
she prepares a sequence of $n$ states, randomly chosen from
$\mid\!\!\Psi\rangle$ or $\mid\!\!\Psi'\rangle$.
They also agree
on a security parameter $n$.
\\
(2) Bob chooses a vector $B= \left(\theta_{1},\theta' _{1},
\phi_{1},\phi'_{1}, \dots,
\theta_{n},\theta' _{n},
\phi_{n},\phi' _{n}\right)$ such that
$\theta_{i}, \theta'_{i}$, $\phi_{i}$, and 
$\phi'_{i}$ satisfy one of the following
relations:
\\
$(1)$  $\theta_{i}=\theta' _{i} = 90^{\circ}$ and
$\phi_{i}+\phi' _{i} = 0^{\circ}$,
\\
$(2)$ $\theta_{i}=\theta' _{i}$ and
$\phi_{i}+\phi' _{i} = 0^{\circ}$,
\\
$(3)$ $\theta_{i}+\theta' _{i} = 180^{\circ}$ and
$\phi_{i}+\phi' _{i} = 0^{\circ}$,
\\
$(4)$ $\theta_{i}=\theta' _{i} = 90^{\circ}$ and
$\phi_{i}+\phi'_{i} = 90^{\circ}$,
\\
$(5)$ $\theta_{i}=\theta' _{i} $ and
$\phi_{i}+\phi'_{i} = 90^{\circ}$,
\\
$(6)$ $\theta_{i}+\theta' _{i}=180^{\circ}$ and
$\phi_{i}+\phi' _{i} = 90^{\circ}$.
\\
Bob will measure the spin of the first (second) particle at polar angle
$\theta_{i}(\theta' _{i})$ and azimuthal angel 
$\phi_{i}(\phi'_{i})$. He
keeps the vector $B$ secret.
\\
(4)\(\mathop{\bf Do}\limits_{i=1}^{n}\)
Alice sends the $i$th EPR pair to Bob. Bob measures the spin of
the first
particle, $m_{1}^{a}$, along axis $\vec{a}$
at polar (azimuthal) angle $\theta_i (\phi _i)$
and spin of the second particle, $m_{1}^{b}$,
along axis $\vec{b}$
at polar (azimuthal) angle
$\theta'_i (\phi'_i)$.
Bob keeps the results of his measurements secret.

Note that Bob does not learn any information about
the bit $\lambda$
since if Alice selects states
$\mid\!\Phi\rangle$ or $\mid\!\Phi'\rangle$, then
\begin{eqnarray}
p(m_{1}^{a}m_{1}^{b}=1)&=&\frac{1}{2} \left ( \frac{1 +
E_{\Phi}(\vec{a}, \vec{b})}{2}
+\frac{1 +
E_{\Phi'}(\vec{a}, \vec{b})}{2}
\right),\\ \nonumber
&=& \frac{1+cos\theta_1 cos\theta_2}{2},\\ \nonumber
p(m_{1}^{a}m_{1}^{b}=-1)&=&\frac{1-cos\theta_1 cos\theta_2}{2},
\end{eqnarray}
and if she selects states
$\mid\!\Psi\rangle$ or $\mid\!\Psi'\rangle$, then
\begin{eqnarray}
p(m_{1}^{a}m_{1}^{b}=1)=\frac{1+cos\theta_1 cos\theta_2}{2},\\ \nonumber
p(m_{1}^{a}m_{1}^{b}=-1)=\frac{1-cos\theta_1 cos\theta_2}{2}.
\end{eqnarray}
Thus according to the standard rules of quantum mechanics, Bob does not
learn any information about Alice's bit no matter along which axis
he performs his measurement.

We now consider the opening of the commitment.
Alice and Bob initiate the following steps:
\\
(1) Alice reveals the bit $\lambda$
to Bob [As previously stated, $\lambda = 1$
indicates that Alice has chosen state $\mid\!\Phi\rangle$,
or $\mid\!\Phi'\rangle$,
and  $\lambda = 0$ indicates that Alice has chosen the state
$\mid\!\Psi\rangle$ or $\mid\!\Psi'\rangle$].
\\
(2)
If $\lambda = 1$, then
$\mathop{\bf DO}\limits_{i=1}^{n}$ Bob checks that if the
$i$th EPR state is
$\mid\!\Phi\rangle$, then
$m^{1}_{a}m^{2}_{b}=1$ whenever
$\theta_{i}=\theta' _{i} = 90^{\circ}$ and
$\phi_{i}+\phi' _{i} = 0^{\circ}$,
and whenever
$\theta_{i}=\theta' _{i}$ and
$\phi_{i}+\phi' _{i} = 0^{\circ}$. If the $i$th EPR state is
is $\mid\!\Phi'\rangle$, then Bob checks that
$m^{1}_{a}m^{2}_{b}=-1$ whenever
$\theta_{i}=\theta'_{i} = 90^{\circ}$ and
$\phi_{i}+\phi'_{i} = 0^{\circ}$,
and whenever
$\theta_{i}+\theta' _{i}=180^{\circ}$ and
$\phi_{i}+\phi' _{i} = 0^{\circ}$.
\\
If $\lambda = 0$,
$\mathop{\bf DO}\limits_{i=1}^{n}$ Bob checks that if
the $i$th EPR state is
$\mid\!\Psi\rangle$, then
$m^{1}_{a}m^{2}_{b}=1$ whenever
$\theta_{i}=\theta'_{i} = 90^{\circ}$ and
$\phi_{i}+\phi' _{i} = 90^{\circ}$,
and whenever
$\theta_{i}=\theta' _{i}$ and
$\phi_{i}+\phi' _{i} =90^{\circ}$. If the $i$th EPR state is
is $\mid\!\Psi'\rangle$ then Bob checks that
$m^{1}_{a}m^{2}_{b}=-1$ whenever
$\theta_{i}=\theta'_{i} = 90^{\circ}$ and
$\phi_{i}+\phi'_{i} = 90^{\circ}$, and 
whenever
$\theta_{i}+\theta'  _{i} =180^{\circ}$ and
$\phi_{i}+\phi'_{i} = 90^{\circ}$.
\\
(3) If these conditions are satisfied, then Bob accepts that
Alice had indeed committed to the bit $\lambda$.

We now show that the above bit commitment
protocol is perfectly secure against
cheaters with unlimited computing power
provided Alice does not have the technology to
store
an EPR particle for an arbitrarily long period of time.
First we note that if
Alice is honest, and if no transmission errors
occur \cite{12}, then condition $(2)$ is always satisfied.
Now suppose that a cheating Alice tries to commit
in a way that will enable her to change $\lambda$ at
a later time. In order to achieve this, she must tell
Bob that she had selected (for example) state
$\mid\!\Phi\rangle$,
when in reality she had committed to the state
$\mid\!\Psi\rangle$.
Consider an instance when Bob measures
the spin of the first particle along axis $\vec{a}$ and
the spin of the second particle along axis $\vec{b}$, and
obtains $m_{1}^{a}m_{2}^{b}=1$.
If she cheats and tells Bob
that she had used $\mid\!\Phi\rangle$, then
the probability that her guess is correct is
$\large \frac{1}{2}$.
Therefore, in the long run, the probability that Alice cheats
and succeeds is
$\left(\frac{1}{2}\right)^{\frac{n}{2}}$
[note that Bob uses approximately $\frac{\displaystyle n}
{\displaystyle 2}$ particles
to reach a decision: When Alice tells Bob that she
has committed to state $\mid\!\Phi\rangle$ or $\mid\!\Phi'\rangle$,
then Bob considers only
instances when $\phi_{i} + \phi'_{i} =0^\circ$
which happens for approximately $\frac{\displaystyle n}
{\displaystyle 2}$ particles.
Similarly when Alice tells Bob that she has committed to state
$\mid\!\Psi\rangle$ or $\mid\!\Psi'\rangle$,
then Bob considers only instances when $\phi_{i} + \phi'_{i} =90^\circ$
which again happens for approximately $\frac{\displaystyle n}
{\displaystyle 2}$ particles].
For sufficiently large security parameter $n$, the probability
of success of a cheating Alice can be made arbitrarily small.

We now show that the proposed protocol is not secure
against EPR attack (see also [8-11]), that is, we show
that the protocol is not secure against a cheating Alice who
has the technology to store the third particle of the following
EPR state
$\mid\!A\rangle=\frac{1}{\sqrt{2}}\left (
\mid\uparrow \uparrow  \uparrow\rangle
+\mid\downarrow \downarrow \downarrow \rangle \right)$
for an arbitrarily long
period of time.
We
consider a spin variable at polar angle $\theta$ and at
azimuthal angle $\phi$ along direction $\vec{n}$.
The states of spin-up
$\mid\uparrow\rangle$
and spin-down
$\mid\downarrow\rangle$
can be expanded in terms of
the states of spin-up
$\mid\!+\rangle$ and spin-down 
$\mid\!-\rangle$
along direction $\vec{n}$
so that we have the expressions
\begin{eqnarray}{\nonumber}
\mid\uparrow\rangle=e^{\left(i\phi/2\right)}
\left[ \cos \left(\theta/2\right)\mid\!+\rangle
-\sin \left(\theta/2\right)\mid\!-\rangle \right], \\
\mid\downarrow\rangle=e^{\left( -i\phi/2\right)}
\left[\sin \left(\theta/2\right)\mid\!+\rangle
+\cos  \left(\theta/2\right)\mid\!-\rangle \right].
\end{eqnarray}
By expanding $\mid \uparrow\rangle$
and $\mid \downarrow\rangle$ in terms of
$\mid x\rangle$ and
$\mid -x\rangle$, we obtain
\begin{eqnarray}{\nonumber}
\mid \uparrow\rangle=\frac{1}{\sqrt{2}}\left (
\mid\!x\rangle\,-\mid\!-\,x\rangle \right), \\
\mid\downarrow\rangle=\frac{1}{\sqrt{2}}\left (
\mid\!x\rangle\,+\mid\!-\,x\rangle \right).
\end{eqnarray}
Similarly
by expanding $\mid \uparrow\rangle$
and $\mid \downarrow\rangle$ in terms of
$\mid y\rangle$ and
$\mid -y\rangle$, we obtain
\begin{eqnarray}{\nonumber}
\mid\uparrow\rangle=\frac{(1+i)}{2}\left (
\mid\!y\rangle\,-\mid\!-\,y\rangle \right), \\
\mid\downarrow\rangle=\frac{(1-i)}{2}\left (
\mid\!y\rangle\,+\mid\!-\,y\rangle \right).
\end{eqnarray}

Using Eqs. $(9)$ and
expanding the third particle of state $\mid\!A\rangle$
in terms of
$\mid x\rangle$ and $\mid -x\rangle$,
we obtain
\begin{eqnarray}
\mid\!A\rangle=\frac{1}{2}\left [
\mid\uparrow \uparrow \rangle \left(\mid\!x\rangle\,-\mid\!-\,x\rangle
\right) \right]
+\frac{1}{2}\left [
\mid\downarrow \downarrow \rangle
\left( \mid\!x\rangle\,+\mid\!-\,x\rangle
\right) \right].
\end{eqnarray}
Rearranging the terms,
\begin{eqnarray}
\mid\!A\rangle=\frac{1}{\sqrt{2}}\left [
\mid\!\phi\rangle
\mid\!x\rangle \, -
\mid\!\phi'\rangle 
\mid\!-\,x\rangle \right].
\end{eqnarray}
Similarly by expanding the
third particle of state
$\mid\!A\rangle$ in terms the $\mid\!y\rangle$ and
$\mid\!-\,y\rangle$, we obtain
\begin{eqnarray}
\mid\!A\rangle=\frac{1+i}{2}\left [
\mid\!\psi'\rangle
\mid\!y\rangle \, -
\mid\!\psi\rangle  
\mid\!-\,y\rangle \right].
\end{eqnarray}
Now if Alice wants to pretend that she had committed to bit
$0$, that is, if she wants to pretend
she had chosen states $\mid\!\Phi\rangle$ or
$\mid\!\Phi'\rangle$, then
she measures the spin of the third particle along the $x$ axis. If the
result of her measurement is $1$ ($-1$), then she
tells Bob that she had chosen state $\mid\!\Phi\rangle$
($\mid\!\Phi'\rangle$). Similarly if Alice
wants to pretend that she had committed to bit
$1$, that is, if she wants to pretend that she had chosen states
$\mid\!\Psi\rangle$ or $\mid\!\Psi'\rangle$, then
she measures the spin of the third particle along the $y$ axis. If the
result of her measurement is $1$ ($-1$), then she
tells Bob that she had chosen state $\mid\!\Psi'\rangle$
($\mid\!\Psi\rangle$).

In summary, we
have shown that the present protocol is secure even against
cheaters with unlimited computing power,
However, the proposed scheme is not secure against a cheating Alice
who has the technology to store an EPR particle for an
arbitrarily long period of time.

I am grateful to H. K. Lo for helpful discussions. I also thank
D. Mayers for sending me his unpublished
results on the insecurity of all bit commitment protocols.

\pagebreak


\begin{thebibliography}{99}
\bibitem{12} There is a practical problem
that is very easy to deal with. In practical detectors
and polarizers, transmission errors occur due to
misalignment of polarizers, dark counts, etc.
If the error rate is $\eta\%$ then Bob should check that
condition (2) in opening
the commitment is satisfied 
for at least $(100-\eta)\%$ of the time.
\end{thebibliography}
\end{document}